\documentclass[Physsubmission, Phys]{SciPost}
\hypersetup{
    colorlinks,
    linkcolor={red!50!black},
    citecolor={blue!50!black},
    urlcolor={blue!80!black}
}
\usepackage{slashed}
\usepackage[bitstream-charter]{mathdesign}
\DeclareSymbolFont{usualmathcal}{OMS}{cmsy}{m}{n}
\DeclareSymbolFontAlphabet{\mathcal}{usualmathcal}

\begin{document}

\begin{center}{\Large \textbf{
NLO collinear factorization of large mass  diphoton photoproduction amplitude
}}\end{center}

% TODO: write the author list here. Use initials + surname format.
% Separate subsequent authors by a comma, omit comma at the end of the list.
% Mark the corresponding author with a superscript *.
\begin{center}
O. Grocholski\textsuperscript{1,2$\star$},
B. Pire\textsuperscript{3},
P. Sznajder\textsuperscript{1},
L. Szymanowski\textsuperscript{1},
J. Wagner \textsuperscript{1}.
\end{center}

\begin{center}
{\bf 1} National Centre for Nuclear Research (NCBJ), Pasteura 7, 02-093 Warsaw, Poland
\\
{\bf 2} Institute of Theoretical Physics, Faculty of Physics, University of Warsaw, Pasteura 5, 02-093
Warsaw, Poland
\\
{\bf 3} CPHT, CNRS, École Polytechnique, I.P. Paris, 91128 Palaiseau, France
\\
* o.grocholski@student.uw.edu.pl
\end{center}

\begin{center}
\today
\end{center}

% For convenience during refereeing (optional),
% you can turn on line numbers by uncommenting the next line:
%\linenumbers

\definecolor{palegray}{gray}{0.95}
\begin{center}
\colorbox{palegray}{
  \begin{tabular}{rr}
  \begin{minipage}{0.1\textwidth}
    \includegraphics[width=22mm]{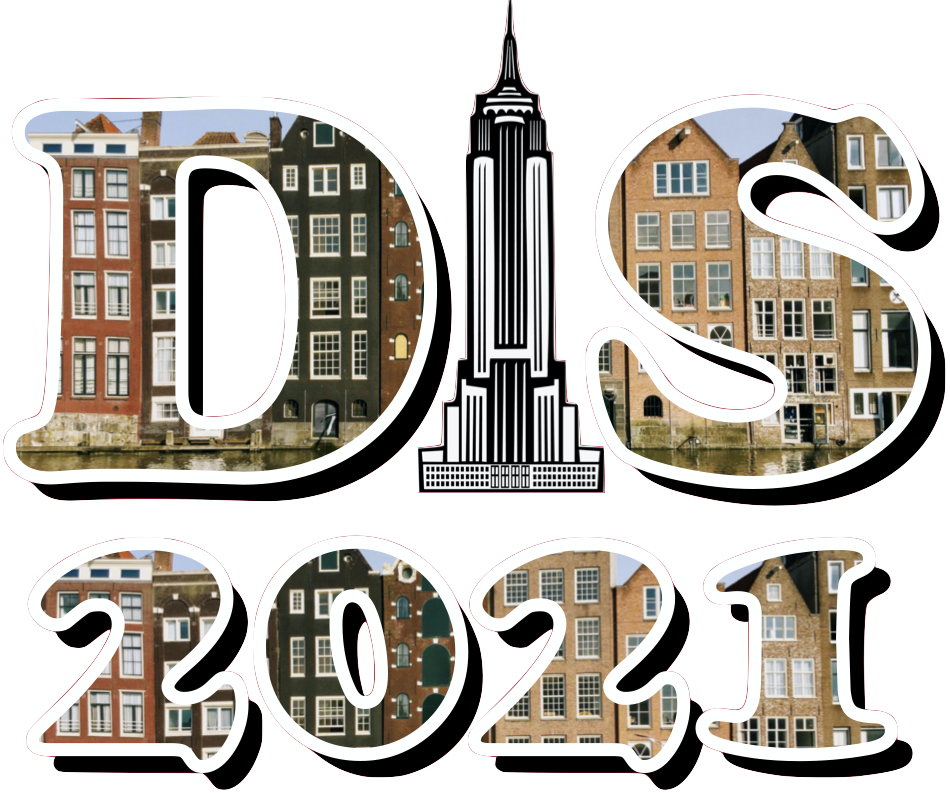}
  \end{minipage}
  &
  \begin{minipage}{0.75\textwidth}
    \begin{center}
    {\it Proceedings for the XXVIII International Workshop\\ on Deep-Inelastic Scattering and
Related Subjects,}\\
    {\it Stony Brook University, New York, USA, 12-16 April 2021} \\
    \doi{10.21468/SciPostPhysProc.?}\\
    \end{center}
  \end{minipage}
\end{tabular}
}
\end{center}

\section*{Abstract}
{\bf
% TODO: write your abstract here.
We calculate large mass diphoton exclusive photoproduction in the framework of collinear QCD factorization at next to leading order in $\alpha_s$ and at leading twist. Collinear divergences of the coefficient function are absorbed by the evolution of the generalized parton distributions (GPDs). This result enlarges the existing factorization proofs to $2 \to  3$ processes, opening new reactions to a trustable extraction of GPDs. 
}

%\vspace{10pt}
%\noindent\rule{\textwidth}{1pt}
%\tableofcontents\thispagestyle{fancy}
%\noindent\rule{\textwidth}{1pt}
%\vspace{10pt}

\section{Introduction}
\label{sec:intro}
Exclusive hard processes such as deeply virtual Compton scattering (DVCS) have been demonstrated to open a window on the tomography of nucleons. After a first rich crop of experimental data~\cite{Kumericki:2016ehc}, it becomes clear~\cite{Bertone:2021yyz} that trustable information on the quark and gluon $3-$dimensional content of hadrons needs to gather experimental data from a set of reactions not restricted to the simplest ones (spacelike, timelike or double DVCS on the one hand, exclusive meson electroproduction on the other hand).   After a Born order study  \cite{Pedrak:2017cpp, Pedrak:2020mfm} which demonstrated its phenomenological interest, we thus studied, at next to leading order (NLO) in the QCD coupling constant $\alpha_s$, the photoproduction of a large mass diphoton on a    nucleon target
\begin{equation}
\gamma(q,\epsilon) + N(p_1,s_1) \rightarrow \gamma(q_1,\epsilon_1) +  \gamma(q_2,\epsilon_2)+ N'(p_2,s_2)\,,
\label{process}
\end{equation}
 in the kinematical domain suitable to a factorization of generalized parton distributions (GPDs) and a hard amplitude, namely large invariant diphoton squared mass $M_{\gamma\gamma}^2 = (q_1+q_2)^2$ and small momentum transfer $-t =-(p_2-p_1)^2 \ll  Q^2$ between the initial and  final nucleons. This reaction, which superficially looks like timelike Compton scattering (TCS) \cite{Mueller:1998fv, Berger:2001xd, Pire:2011st,Moutarde:2013qs} with the dilepton replaced by a diphoton, has a much different structure since the hard amplitude assumed to factorize from the GPDs describes a $2 \to 3$ process, contrarily to the usual reactions  where the coefficient function describes  a $2 \to 2$ process. However, there is yet no all order proof of QCD collinear factorization at leading twist for such processes.
  
 This NLO study is the first step toward the  derivation of collinear factorization of the amplitude of this process. We use dimensional regularization as the main tool of our proof. 
\section{LO results}
 
\begin{figure}[h]
\vspace{-0.5cm}
    \centering
    \includegraphics[width = 0.6 \textwidth]{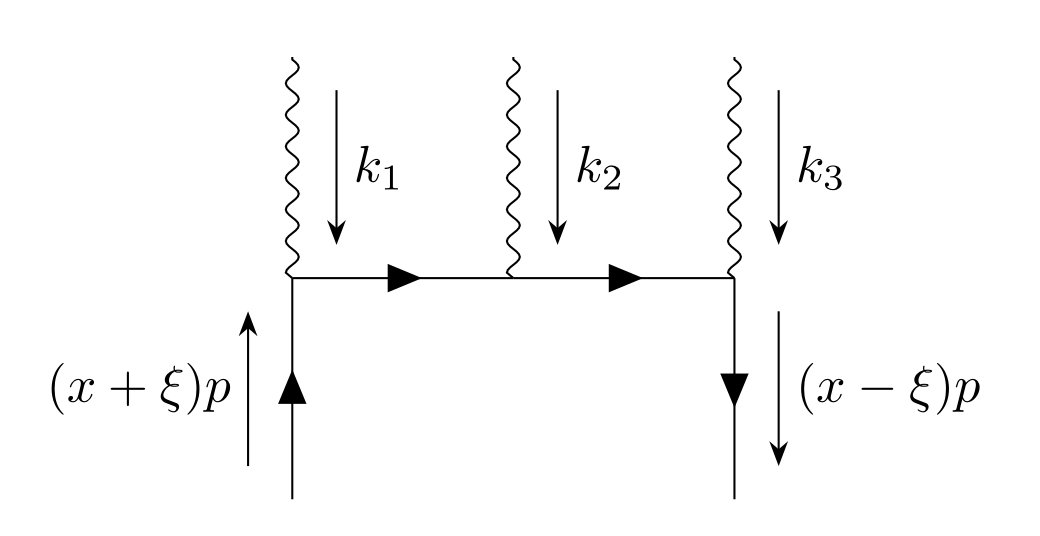}
    \caption{The general form of the LO diagram. As described in the text, to simplify computations we first consider a diagram with 3 incoming photons, and  use afterwards the appropriate substitutions.}
    \label{fig:LO-graph}
\end{figure}
Let us first outline the results of the LO study for photoproduction \cite{Pedrak:2017cpp} and electroproduction \cite{Pedrak:2020mfm}. Due to charge parity constraint, the contributing GPDs  at any order in $\alpha_s$ are the charge-odd quark GPDs (which are not accessed by DVCS) and there is no gluon GPD contribution. The leading order hard vector and axial amplitudes are straightforwardly calculated from the 6 diagrams obtained from Fig. \ref{fig:LO-graph} with various permutations. Their structure is very peculiar since they are proportional to  $\delta(x \pm \xi)$, the skewness parameter $\xi$ being related to Lorentz invariants in a similar way as in the TCS kinematics:
$\xi= \tau / (2-\tau)$ with $\tau = M_{\gamma\gamma}^2/ s$.
To reduce the number of analyzed diagrams, we consider a process with 3 photons with momenta $(k_1,k_2,k_3)$ and sum over all possible permutations.
The vector part of the hard amplitude corresponding to a given permutation reads
\begin{equation}\label{eq-M-LO-0}
	Tr\big[ i\mathcal{M}^0_{1,2,3}~ \slashed{p} \big] = ie_q^3 \frac{1}{2(x+\xi)pk_1 + i0}\frac{1}{-2(x-\xi)pk_3 + i0} Tr \Big( \slashed{\epsilon}_3 ( 
%	\underline{\slashed{p}} 
	- \slashed{k}_3 ) \slashed{\epsilon}_2  
%	( \bar{\slashed{p}} + 
\slashed{k}_1 
%) 
\slashed{\epsilon}_1 \slashed{p} \Big),
\end{equation}
%where $\underline{p} = (x-\xi)p$, $\overline{p} = (x+\xi)p$, 
leading after summing the permutations to:
\begin{eqnarray}
\label{eq-Cq0-recap}
%\begin{aligned}
	&&
	\sum_{\mathrm{perm.}} Tr\Big[ i\mathcal{M}^0_{1,2,3} \slashed{p} \Big]
	%=
	%\mathcal{C}^q_0 (x, \xi, \dots) 
	%
	= 4\frac{ie_q^3}{ s \alpha \bar{\alpha} } \: \mathrm{Im} \Big( \frac{1}{x+\xi - i0} \frac{1}{x-\xi +i0} \Big)\times \\&&  \bigg[(\alpha - \bar{\alpha} )\big( \vec{\epsilon^*}_t(\mathbf{q}_1) \vec{\epsilon^*}_t(\mathbf{q}_2) \big) \big( \vec{p}_t \vec{\epsilon}_t (\mathbf{q}) \big)  - \big(  \vec{p}_t \vec{\epsilon^*}_t(\mathbf{q}_1) \big) \big(  \vec{\epsilon}_t (\mathbf{q}) \vec{\epsilon^*}_t(\mathbf{q}_2) \big) + \big(  \vec{p}_t \vec{\epsilon^*}_t(\mathbf{q}_2) \big) \big(  \vec{\epsilon}_t(\mathbf{q}) \vec{\epsilon^*}_t(\mathbf{q}_1) \big) \bigg] \nonumber \,,
%\end{aligned}
\end{eqnarray}
where $\alpha = 1- \bar \alpha =  \frac{u'}{u' + t'}$, with $t' = (q_1 - q)^2 = t - M_{\gamma\gamma}^2 - u'$.

The axial LO amplitude corresponding to a given permutation of photons reads:
\begin{equation}
    Tr\Big[ i\mathcal{M}^0_{1,2,3} \gamma^5 \slashed{p} \Big] = 
%    ie_q^3 \frac{s^{-2}}{\beta_1 \beta_3} \frac{1}{x+\xi + i0_1} \frac{1}{x-\xi -i0_3} 
    ie_q^3 \frac{1}{2(x+\xi)pk_1 + i0}\frac{1}{-2(x-\xi)pk_3 + i0} 
    Tr\big( \slashed{\epsilon}_3 (-\slashed{k}_3) \slashed{\epsilon}_2 \slashed{k}_1 \slashed{\epsilon}_1 \gamma^5 \slashed{p} \big).
\end{equation}
%where $\beta_i = 2pk_i$ and $0_i = 0\times \textrm{sgn} \beta_i$. 
%The exact form of the trace is not important here. Note, that the exchange $1\leftrightarrow 3$ changes the sign of the trace, and switches the infinitesimal parts $i0_1 \leftrightarrow i0_3$. 
After summation over all permutations of photons we get:
%, only the imaginary part of amplitudes corresponding to $(k_1, k_2, k_3) = (q, -q_{1/2} -q_{2/1})$ and $(k_1, k_2, k_3) = (-q_{1/2} -q_{2/1}, q)$ survives, so that
\begin{equation}\begin{aligned}
    &\sum_{\mathrm{perm.}} Tr\Big[ i\mathcal{M}^0_{1,2,3} \gamma^5 \slashed{p} \Big] = -2e_q^3 s^{-2} \:  \mathrm{Im} \: \Big[ \frac{1}{x+\xi - i0} \frac{1}{x-\xi -i0} \Big] \times  \\ & \Big( \frac{1}{\alpha} Tr\big( \slashed{\epsilon}(q) \slashed{q} \slashed{\epsilon}^*(q_2) \slashed{q}_1 \slashed{\epsilon}^*(q_1) \gamma^5 \slashed{p} \big) + \frac{1}{\bar{\alpha}} Tr\big( \slashed{\epsilon}(q) \slashed{q} \slashed{\epsilon}^*(q_2) \slashed{q}_1 \slashed{\epsilon}^*(q_1) \gamma^5 \slashed{p} \big) \Big) 
%    \\
%    & \equiv -2e_q^3 s^{-2} \Tilde{\mathcal{A}} \: \mathrm{Im} \: \Big[ \frac{1}{x+\xi - i0} \frac{1}{x-\xi -i0} \Big].
    \end{aligned}
\end{equation}
\begin{figure}[h]
    \centering
    \includegraphics[width = 0.6 \textwidth]{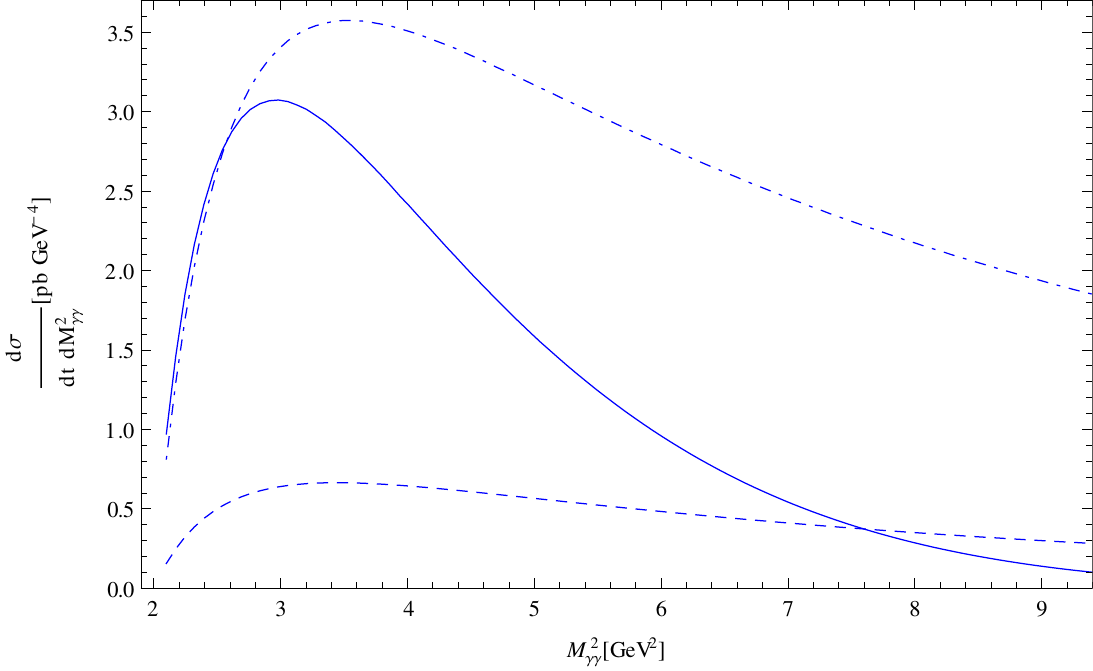}
    \caption{The $M_{\gamma\gamma}^2$ dependence of the LO cross-section for the process $\gamma p \to \gamma \gamma p'$ at $s = 20$  (resp. $100$, resp $10^6$) GeV$^2$ : full (resp. dashed, resp. dash-dotted) curve. The highest energy curve is multiplied by $10^5$.}
    \label{fig:LO}
\end{figure}
The detailed phenomenological study  performed in \cite{Pedrak:2020mfm} and exemplified in Fig. \ref{fig:LO} showed that  the cross-section is dominated by the contribution of vector GPDs, and that the cross-sections are large enough for a promising phenomenological study in quasi real photoproduction at electron machines, mostly at medium energy because of the zero contribution of gluons and sea-quarks. 
\section{The NLO amplitude}

\subsection{Toward a factorization proof of the amplitude}
As emphasized above, no factorization proof exists for any exclusive process with a genuine $2 \to 3$ hard amplitude. The first step in establishing such a proof consists in a careful analysis of first order QCD corrections to the Born amplitude. The calculation of such corrections necessarily introduces divergent integrals which we choose to calculate using dimensional regularization. Meanwhile, a consistent NLO calculation of the whole amplitude also needs to take into account the difference between bare and renormalized GPDs. The signature of collinear factorization at this order is the fact that the divergent part of the coefficient function convoluted with the bare GPD leads to a renormalized GPD which obeys the known QCD evolution equations. This is the way we proceed.  
 \subsection{Analysis of the divergent parts in the NLO amplitude}
 
 \begin{figure}[h]
    \centering
    \includegraphics[width = 0.4 \textwidth]{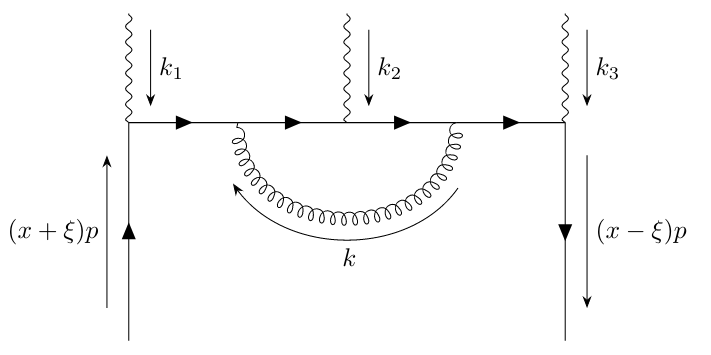}~~~~~
    \includegraphics[width = 0.4 \textwidth]{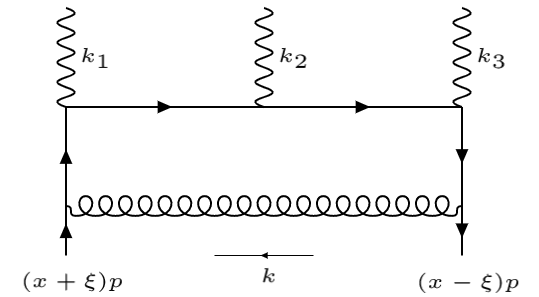}
    \caption{Some NLO diagrams. }
    \label{fig:NLO}
\end{figure}

 This is not the place to write down the technical details of our calculation which will be shortly submitted for publication. The calculation of the NLO diagrams such as those of Fig. \ref{fig:NLO} is straightforward and takes profit from the usual tools developed for similar QCD calculations. 
 
 Our calculation of the divergent part of the vector amplitude yields:
%\begin{eqnarray}\label{eq-factorization-Mcoll}
%\begin{aligned}
%	&&Tr\big[ i\mathcal{M}_{1,2,3}~\slashed{p}\big]_{\mathrm{div.}} = \frac{2}{\varepsilon} \cdot i\frac{\alpha_S}{4\pi}C_F e_q^3\mathcal{A}_{1,2,3} \frac{4 s^{-2}}{\beta_1 \beta_3} \frac{1}{x-\xi-i0\cdot sgn(\beta_3)}\frac{1}{x+\xi+i0\cdot sgn(\beta_1)} \times \nonumber\\
%&& \bigg( 3+ \frac{x+\xi}{\xi}\log \Big{(} \frac{1}{2\xi}\big{(}\xi - x + i0\cdot sgn(\beta_3)\big{)} \Big{)} - \frac{x-\xi}{\xi}\log \Big{(} \frac{1}{2\xi}\big{(}\xi + x + i0\cdot sgn(\beta_1)\big{)} \Big{)} \bigg)\,,
%\end{aligned}
%\end{eqnarray}
%which can also be written as
\begin{equation}\label{eq-Cqcoll-fin}
\begin{aligned}
&\sum_{\mathrm{perm.}} Tr\big[ i\mathcal{M}_{1,2,3}~\slashed{p}\big]_{\mathrm{div.}} = \frac{2}{\varepsilon} \cdot 4\frac{e_q^3}{ s \alpha \bar{\alpha} } \frac{\alpha_S}{4\pi}C_F \bigg[ (\alpha - \bar{\alpha} )\big( \vec{\epsilon^*}_t(\mathbf{q}_1) \vec{\epsilon^*}_t(\mathbf{q}_2) \big) \big( \vec{p}_t \vec{\epsilon}_t (\mathbf{q}) \big) + \\& - \big(  \vec{p}_t \vec{\epsilon^*}_t(\mathbf{q}_1) \big) \big(  \vec{\epsilon}_t (\mathbf{q}) \vec{\epsilon^*}_t(\mathbf{q}_2) \big) + \big(  \vec{p}_t \vec{\epsilon^*}_t(\mathbf{q}_2) \big) \big(  \vec{\epsilon}_t(\mathbf{q}) \vec{\epsilon^*}_t(\mathbf{q}_1) \big) \bigg]\times \\
& \: \mathrm{Im} \bigg[ \frac{1}{x-\xi+i0}\frac{1}{x+\xi-i0} \bigg( 3+ \frac{x+\xi}{\xi}\log \Big{(} \frac{1}{2\xi}\big{(}\xi - x - i0 \big{)} \Big{)} - \frac{x-\xi}{\xi}\log \Big{(} \frac{1}{2\xi}\big{(}\xi + x - i0\big{)} \Big{)} \bigg) \bigg].
\end{aligned}
\end{equation}
%\begin{eqnarray}
%\begin{aligned}
%	&&Tr\big[ i\mathcal{M}_{1,2,3}~\slashed{p}\big]_{\mathrm{div.}} = -\frac{2}{\varepsilon} \cdot \frac{\alpha_S}{4\pi}C_F \cdot Tr\big[ i\mathcal{M}^0_{1,2,3}~\slashed{p}\big]  \times  \\
%&& \bigg( 3+ \frac{x+\xi}{\xi}\log \Big{(} \frac{1}{2\xi}\big{(}\xi - x + i0\cdot sgn(\beta_3)\big{)} \Big{)} - \frac{x-\xi}{\xi}\log \Big{(} \frac{1}{2\xi}\big{(}\xi + x + i0\cdot sgn(\beta_1)\big{)} \Big{)} \bigg)\nonumber \,.
%\end{aligned}
%\end{eqnarray}
%The term $\mathcal{C}^q_{coll.}$ is defined by
%\begin{equation}
% \sum_{\mathrm{permutations}} Tr\big[ %\mathcal{M}_{1,2,3}~\slashed{p}\big]_{\mathrm{div.}} = -\frac{2}{\varepsilon}\: %\mathcal{C}^q_{coll.}.
%\end{equation}
To prove factorization at this order, we check that
\begin{eqnarray}\label{eq-to-verify}
%\begin{aligned}
&&\int_{-1}^{1} dy\: K^{qq}(y,x) \mathrm{Im} \Big( \frac{1}{y+\xi - i0} \frac{1}{y-\xi +i0} \Big) %\: \overset{?}{=} 
=\\
&&\frac{\alpha_S}{4\pi} C_F \: \mathrm{Im} \: \frac{1}{x-\xi+i0}\frac{1}{x+\xi-i0} \bigg( 3+ \frac{x+\xi}{\xi}\log \Big{(} \frac{1}{2\xi}\big{(}\xi - x - i0 \big{)} \Big{)} - \frac{x-\xi}{\xi}\log \Big{(} \frac{1}{2\xi}\big{(}\xi + x - i0\big{)} \Big{)} \bigg)\nonumber \,,
%\end{aligned}
\end{eqnarray}
where the non-singlet quark kernel $K^{qq}$ (the only kernel relevant to our process) reads:
\begin{equation}
	K^{qq}(x,x') = \frac{1}{|\xi|} \frac{\alpha_S}{4\pi} C_F \Bigg[ \rho\Big(\frac{x}{\xi}, \frac{x'}{\xi} \Big) \bigg\{ \frac{x+\xi}{x'+\xi}\Big(1+\frac{2\xi}{x'-x} \Big) \bigg\} + \rho\Big(-\frac{x}{\xi}, -\frac{x'}{\xi} \Big) \bigg\{ \frac{-x+\xi}{-x'+\xi}\Big(1+\frac{2\xi}{x-x'} \Big) \bigg\} \Bigg]_+, \nonumber
\end{equation}
with
\begin{equation}
	\rho\Big(\frac{x}{\xi}, \frac{x'}{\xi} \Big) = \theta\Big(\frac{x}{\xi} - \frac{x'}{\xi} \Big)\theta\Big(\frac{x'}{\xi} + 1 \Big) - \theta\Big( \frac{x'}{\xi}- \frac{x}{\xi}\Big)\theta\Big( -1 -\frac{x'}{\xi}\Big)\,, \nonumber
\end{equation}
where $\theta$ is the step function, and 
%\begin{equation}
$	\int dx \Big[ f(x,x') \Big]_+ g(x) = \int dx f(x,x') \big( g(x) - g(x') \big).$
%\end{equation}
This completes the NLO proof of perturbative factorization of the vector part of the amplitude of our process.
In a similar way, the divergent axial part is found to be:
\begin{eqnarray}
  %  \begin{aligned}
  	&&\sum_{\mathrm{perm.}} Tr\big[ i\mathcal{M}_{1,2,3}~\gamma^5\slashed{p}\big]_{\mathrm{div.}} =
%       &&\mathcal{C}^{q,A}_{\mathrm{coll.}} =
        \frac{2}{\varepsilon} \cdot 2i\frac{\alpha_S}{4\pi}C_F e_q^3 s^{-2} \times \nonumber \\ &&\Big( \frac{1}{\alpha} Tr\big( \slashed{\epsilon}(q) \slashed{q} \slashed{\epsilon}^*(q_2) \slashed{q}_1 \slashed{\epsilon}^*(q_1) \gamma^5 \slashed{p} \big) + \frac{1}{\bar{\alpha}} Tr\big( \slashed{\epsilon}(q) \slashed{q} \slashed{\epsilon}^*(q_2) \slashed{q}_1 \slashed{\epsilon}^*(q_1) \gamma^5 \slashed{p} \big) \Big) \,\times \\
        &&  \mathrm{Im} \bigg[ \frac{1}{x-\xi-i0}\frac{1}{x+\xi-i0} \bigg( 3+ \frac{x+\xi}{\xi}\log \Big{(} \frac{1}{2\xi}\big{(}\xi - x + i0 \big{)} \Big{)} - \frac{x-\xi}{\xi}\log \Big{(} \frac{1}{2\xi}\big{(}\xi + x - i0\big{)} \Big{)} \bigg) \bigg]\nonumber \,.
   % \end{aligned}
\end{eqnarray}
As in the vector part, one verifies the factorization by explicit computation.

  \section{Conclusions}
  A complete proof of QCD collinear factorization for the amplitude of process (\ref{process}) requires some further work. The first step we accomplished allows us to be optimistic with regards to its completion.  The numerical consequences of our NLO description of process (\ref{process}) are under progress, both for reactions at electron machines at medium (JLab) or high energy \cite{AbdulKhalek:2021gbh,Anderle:2021wcy} and in ultraperipheral reactions at LHC~\cite{Pire:2008ea}.
  
  \paragraph*{Acknowledgements.}
  
\noindent
 The work of O.G. is financed by the budget for science in 2020-2021, as a research project under the "Diamond Grant" program. The works of J.W. and L.S. are respectively supported by the grants 2017/26/M/ST2/01074 and 2019/33/B/ST2/02588 of the National Science Center in Poland. This work is partly supported by the Polish-French collaboration agreements Polonium, by the Polish National Agency for Academic Exchange and COPIN-IN2P3 and by the European Union’s Horizon 2020 research and innovation programme under grant agreement No 824093.  

\bibliography{diphoton.bib}

\nolinenumbers

\end{document}